\documentclass[osajnl,preprint,showpacs,superscriptaddress,12pt]{revtex4-1} 

\usepackage{amsmath,amssymb,amsfonts}
\usepackage{graphicx}
\usepackage{color}

\begin{document}

\title{Gilmore-Perelomov symmetry based approach to photonic lattices}

\author{Liliana Villanueva Vergara and B. M. Rodr\'{\i}guez-Lara$^{\ast}$}

\address{Instituto Nacional de Astrof\'{\i}sica, \'Optica y Electr\'onica, Calle Luis Enrique Erro No. 1, Sta. Ma. Tonantzintla, Pue. CP 72840, M\'exico.}

\email{$^{\ast}$bmlara@inaoep.mx}


\begin{abstract}
We revisit electromagnetic field propagation through tight-binding arrays of coupled photonic waveguides,  with properties independent of the propagation distance, and recast it as a symmetry problem.
We focus our analysis on photonic lattices with underlying symmetries given by three well-known groups, $SU(2)$, $SU(1,1)$ and Heisenberg-Weyl, to show that disperssion relations, normal states and impulse functions can be constructed following a Gilmore-Perelomov coherent state approach.
Furthermore, this symmetry based approach can be followed for each an every lattice with an underlying symmetry given by a dynamical group. \\
\end{abstract}


\maketitle

%
\section{Introduction}

Photonic lattices, that is, tight-binding arrays of photonic waveguides, have proved a valuable tool in the classical simulation of quantum physics \cite{Longhi2009p243,Longhi2011p453}.
Vice versa, the insight gained from these quantum-classical analogies has been useful as a tool for photonic lattice design \cite{RodriguezLara2015p}.
This comes from the fact that the differential equation set describing the propagation of field amplitudes through an array of nearest neighbor coupled waveguides \cite{Jones1965p261,Snyder1972p1267,Kogelnik1972p2327,Yariv1973p919,Streifer1987p1,Haus1987p16,Huang1994p963},
\begin{eqnarray}
-i \frac{d}{dz} \mathcal{E}_{l}(z) &=&  \omega f(l,z) \mathcal{E}_{l}(z) +  \lambda \left[ g(l,z)\mathcal{E}_{l-1}(z) + g(l+1,z) \mathcal{E}_{l+1}(z) \right], 
\end{eqnarray}
with $l=0,1,2,\ldots$ and the restriction $g(0,z)=0$, can be cast in a vector differential equation form similar to  Schr\"odinger equation,
\begin{eqnarray}
-i \frac{d}{dz} \vert \mathcal{E}(z) \rangle &=&  \hat{H}(z) \vert \mathcal{E}(z) \rangle. \label{eq:SchEq}
\end{eqnarray}
Here, the field at the $l$-th waveguide is $\mathcal{E}_{l}(z)$,  the effective waveguide refractive index is $\omega f(l,z)$, and the effective nearest neighbor coupling $\lambda g(l,z)$  \cite{RodriguezLara2015p}. 
We have borrowed Dirac notation from quantum mechanics, such that vectors are written as kets,
\begin{eqnarray}
\vert \mathcal{E}(z) \rangle &=& \left( \begin{array}{c} \mathcal{E}_{0}(z) \\ \mathcal{E}_{1}(z) \\ \mathcal{E}_{2}(z) \\ \vdots \end{array} \right), \\
 &\equiv& \sum_{l=0} \mathcal{E}_{l}(z) \vert j \rangle,
\end{eqnarray} 
that collect all the information from the field amplitudes, and the real-symmetric tridiagonal matrices that collect all the properties of the tight binding array are written as operators,
\begin{eqnarray}
\hat{H}(z) &=& \left( \begin{array}{ccccc}
\omega f(0,z)  & \lambda g(0,z) & 0 & \ldots  & \ldots\\
\lambda g(0,z) & \omega f(1,z) & \lambda g(1,z) & \ldots &  \ldots \\
0 & \lambda g(1,z) & \omega f(2,z) & \lambda g(2,z) &  \ldots \\
\vdots & \vdots & \ddots &\ddots &\ddots
\end{array}\right), \\
&\equiv& \omega \hat{A}_{0}(z) + \lambda \left[ \hat{A}_{+}(z) + \hat{A}_{-}(z) \right].
\end{eqnarray}
The operator $\hat{A}_{0}(z) = f(\hat{n},z)$ is a diagonal matrix, the operators $\hat{A}_{+}(z) = g(\hat{n},z) \hat{V}^{\dagger}$ and $\hat{A}_{-}(z) = \hat{V} g(\hat{n},z)$ are lower- and upper-diagonal matrices in terms of the step, $\hat{n} \vert j \rangle = j \vert j \rangle$, up-step, $\hat{V}^{\dagger} \vert j \rangle = \vert j+1 \rangle$, and down-step, $\hat{V} \vert j \rangle = \vert j-1 \rangle$ operators.
Note that the functions related to the effective index and coupling parameters, $f(l,z)$ and $g(l,z)$, have been rewritten in terms of the step operator, with action $f(\hat{n},z) \vert l \rangle = f(l,z) \vert l \rangle$ and $g(\hat{n},z)\vert l \rangle = g(l,z) \vert l \rangle$, such that it is straightforward to recover the original differential set.
At this point, it is clear that the field amplitude distribution at a point $z$ in the photonic lattices is given by the solution to the Schr\"odinger-like equation in (\ref{eq:SchEq}),
\begin{eqnarray}
\vert \mathcal{E}(z) \rangle = e^{i \int_{0}^{z} \hat{H}(x) dx} \vert \mathcal{E}(0) \rangle,
\end{eqnarray} 
where the information from the initial field amplitudes distribution impinging the lattice is collected in the vector $\vert \mathcal{E}(0) \rangle$.
It is straightforward to connect this result with the impulse function, $\mathcal{I}_{m,n}(\omega, \lambda, z)$, that provides the field amplitude at the $n$-th waveguide given that the initial field impinged just the $m$-th waveguide, 
\begin{eqnarray}
\mathcal{I}_{m,n}(\omega,\lambda,z) = \langle n \vert  e^{i \int_{0}^{z} \hat{H}(x) dx} \vert m \rangle.
\end{eqnarray}
Note that it is typically hard to calculate the propagator $e^{i \int_{0}^{z} \hat{H}(x) dx}$ but underlying symmetries may provide help \cite{RodriguezLara2014p2083,RodriguezLara2014p013802}. 

Our goal here is to bring forward the fact that underlying symmetries may simplify greatly the calculations of dispersion relations, normal modes and impulse functions for photonic lattices.
We will focus in the simplest case of lattices which parameters do not depend on the propagation distance. 
In this case, if the operators $\hat{A}_{0}$ and $\hat{A}_{\pm}$ form a closed group under commutation $\left[\hat{A}, \hat{B} \right] = \hat{A} \hat{B} - \hat{B} \hat{A}$, it is possible to use so-called disentangling formulas \cite{Wilcox1967p962,Wodkiewicz1985p458,Ban1993p1347} to calculate the corresponding impulse function,
\begin{eqnarray}
\mathcal{I}_{m,n}(\omega,\lambda,z) &=& \langle n \vert  e^{i z \hat{H}} \vert m \rangle, \\
&=&  \langle n \vert  e^{i z \left[\omega \hat{A}_{0} + \lambda \left( \hat{A}_{+} + \hat{A}_{-} \right) \right]} \vert m \rangle, \\
&=& \langle n \vert  e^{a_{+}(z) \hat{A}_{+}} e^{a_{0}(z) \hat{A}_{0}} e^{a_{-}(z) \hat{A}_{-}} \vert m \rangle,
\end{eqnarray}
in terms of a particular class of generalized Gilmore-Perelomov coherent states \cite{Gilmore1972p391,Perelomov1972p222,Zhang1990p867}, 
\begin{eqnarray}
\hat{D}\left[a_{0}(z), a_{\pm}(z)\right] \vert m \rangle &\equiv& e^{a_{+}(z) \hat{A}_{+}} e^{a_{0}(z) \hat{A}_{0}} e^{a_{-}(z) \hat{A}_{-}} \vert m \rangle.
\end{eqnarray}
Thus, these photonic lattices can be seen as optical simulators of generalized Gilmore-Perelomov coherent states.
In the following sections, we will provide three working examples of lattices with an underlying symmetry and show that a symmetry based, Gilmore-Perelomov generalized coherent state approach can help us providing dispersion relations, normal modes and impulse functions.
First we will work with finite lattices with underlying $SU(2)$ symmetry; the so-called $J_{x}$ photonic lattices used to produce coherent transfer between input and output waveguides \cite{Perez-Leija2013p022303,PerezLeija2013p012309} are a particular example of this underlying symmetry.
Then we will move to semi-infinite lattices described by the $SU(1,1)$ group which have as particular realization in the literature the so-called Jacobi lattices \cite{ZunigaSegundo2014p987}. 
Finally, we will explore lattices with underlying Heisenberg-Weyl symmetry that includes as a specific example the so-called Glauber-Fock lattices \cite{PerezLeija2010p2409,Keil2011p103601,RodriguezLara2011p053845,Keil2012p3801,PerezLeija2012p013848} and close with a summary.
In all the following cases, the closed forms for the respective impulse functions, in terms of the corresponding generalized Gilmore-Perelomov coherent states, are new in both the optics and quantum mechanics literature, to the best of our knowledge.
In order to corroborate them, all the analytic results were compared with numeric diagonalization, for normal modes, and with results from a normal-modes-decomposition and small-step propagation, for the impulse functions. 
In all cases analytic and numeric results are in good agreement.

\section{Lattices with underlying $SU(2)$ symmetry.}

\begin{figure}[ht]
\center\includegraphics[scale=1]{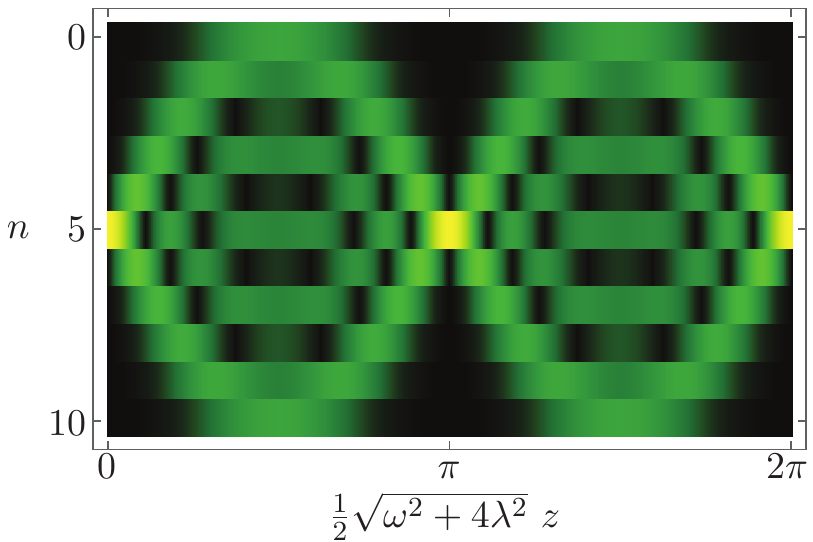}
\caption{Intensity at the $n$-th waveguide, $\vert \mathcal{I}_{m,n}^{(SU(2))}(\omega, \lambda, j, z) \vert^{2}$, for light initially impinging the middle waveguide, $m=5$, of an $SU(2)$ array with parameters $\lambda = 0.4 \omega$ and $j=5$.}
\label{fig:Fig1}
\end{figure}

Let us start with a finite array of coupled waveguides described by the matrix,
\begin{eqnarray}
\hat{H}_{SU(2)}(\omega, \lambda,j) &=& \omega \left( \hat{n} - j \right) + \lambda \left[ \hat{V} \sqrt{ \hat{n} (2j+1 - \hat{n}) } +   \sqrt{ \hat{n} (2j+1 - \hat{n}) } \hat{V}^{\dagger}  \right],\\
&\equiv& \omega \hat{J}_{0} + \lambda ( \hat{J}_{-} + \hat{J}_{+} ),
\end{eqnarray}
in terms of the $SU(2)$ group $[\hat{J}_{+}, \hat{J}_{-}]=2 \hat{J}_{0}$ and $[\hat{J}_{0}, \hat{J}^{\pm}]= \pm \hat{J}_{\pm}$ \cite{Barut1980} and $j$ a positive integer such that $2j+1$ is the size of the lattice.
It is straightforward to construct a Gilmore-Perelomov displacement operator \cite{Gilmore1972p391,Perelomov1972p222,Zhang1990p867},
\begin{eqnarray}
\hat{D}_{SU(2)}(\theta) &=& e^{ \theta \hat{J}_{+} - \theta^{\ast} \hat{J}_{-} }, \\
&=&  e^{\frac{\theta}{\vert \theta \vert} \tan \vert \theta \vert \hat{J}_{+}} e^{\ln \sec^{2} \vert \theta \vert \hat{J}_{0}} e^{- \frac{\theta^{\ast}}{\vert \theta \vert} \tan \vert \theta \vert \hat{J}_{-}},
\end{eqnarray}
that diagonalizes this matrix,
\begin{eqnarray}
\hat{D}_{SU(2)}(\theta) \hat{H}_{SU(2)}(\omega, \lambda) \hat{D}_{SU(2)}^{\dagger}(\theta) = \sqrt{\omega^{2} + 4 \lambda^{2}} ~\left( \hat{n} - j \right) , 
\end{eqnarray}
with 
\begin{eqnarray}
\tan 2 \theta= \frac{2 \lambda}{\omega}.
\end{eqnarray}
Thus, we can immediately construct the dispersion relation for this photonic lattice,
\begin{eqnarray}
\Omega_{SU(2)}(m, \omega, \lambda,j) &=& \sqrt{\omega^{2} + 4 \lambda^{2}}  \left( m - j \right),
\end{eqnarray}
and its normal modes, given by what we are going to call $SU(2)$ displaced number states, such that the $n$-th component of the $m$-th normal mode is given by the expression,
\begin{eqnarray}
\langle n \vert \Omega_{SU(2)}(m,\omega,\lambda,j) \rangle &\equiv& \langle n  \vert \hat{D}_{SU(2)}(-\theta) \vert m \rangle, \\
&=& (-1)^n \sqrt{ \left( \begin{array}{c} 2j \\ m \end{array}\right) \left( \begin{array}{c} 2j \\ n \end{array}\right)} \left( \cos \vert \theta \vert \right)^{2j-m-n} ( \sin \vert \theta \vert)^{m+n} \times \nonumber \\
&& \times  K_{m}\left(n,\sin^{2} \vert \theta \vert , 2j  \right), 
\end{eqnarray}
where we have used the fact that the lattice parameters are real and positive, $\lambda>0$ and $\omega \ge 0$. 
The notation $\left(\begin{array}{c} a \\ b \end{array} \right)$ and $K_{n}(x,p,N)$ stand for the binomial coefficient and Krawtchouk polynomials \cite{Askey1975}, which are well known in the discrete optics community \cite{Wolf2007p651,Vicent2008p1875,Wolf2009p509,Vicent2011p808}, in that order.
Now, it is also possible to compute the impulse function of this array by using standard disentangling formulas for $SU(2)$ \cite{Wilcox1967p962,Wodkiewicz1985p458,Ban1993p1347}, and some algebra, in terms of generalized Gilmore-Perelomov coherent states,
\begin{eqnarray}
\mathcal{I}_{m,n}^{(SU(2))}(\omega, \lambda, j, z)&=&   \langle n  \vert e^{i z \hat{H}_{SU(2)}} \vert m \rangle, \\
&=& \sqrt{ \left( \begin{array}{c} 2j \\ m \end{array}\right) \left( \begin{array}{c} 2j \\ n \end{array}\right)} \frac{\left(2 i \lambda \sin \frac{z}{2} \sqrt{\omega^{2} + 4 \lambda^{2}}  \right)^{m+n}}{\left(\omega^{2} + 4 \lambda^{2}\right)^{j}} \times \nonumber \\
&& \times 
\left(\sqrt{\omega^{2} + 4 \lambda^{2}} \cos \frac{z}{2} \sqrt{\omega^{2} + 4 \lambda^{2}} - i \omega \sin \frac{z}{2}\sqrt{\omega^{2} + 4 \lambda^{2}} \right)^{2j-m-n} \times \nonumber \\
&& \times  K_{m}\left(n, \frac{ 4 \lambda^{2}   }{\omega^{2} + 4 \lambda^{2}}\sin^{2}\frac{z}{2}\sqrt{\omega^{2} + 4 \lambda^{2}} , 2j  \right) .
\end{eqnarray}
Note that the impulse function will show a $2 \pi / \sqrt{\omega^{2} + 4 \lambda^{2}}$ periodicity due to the squared sine argument in the Krawtchouk polynomials.
Figure \ref{fig:Fig1} shows the periodicity discussed above in the intensity, that is, the squared impulse function, $\vert \mathcal{I}_{m,n}^{(SU(2))}(\omega, \lambda, j, z) \vert^{2}$, of an initial field amplitude impinging just the central waveguide of an $SU(2)$ lattice with parameters $\lambda=0.4 \omega$ and $j=5$.

\begin{figure}[ht]
\center\includegraphics[scale=1]{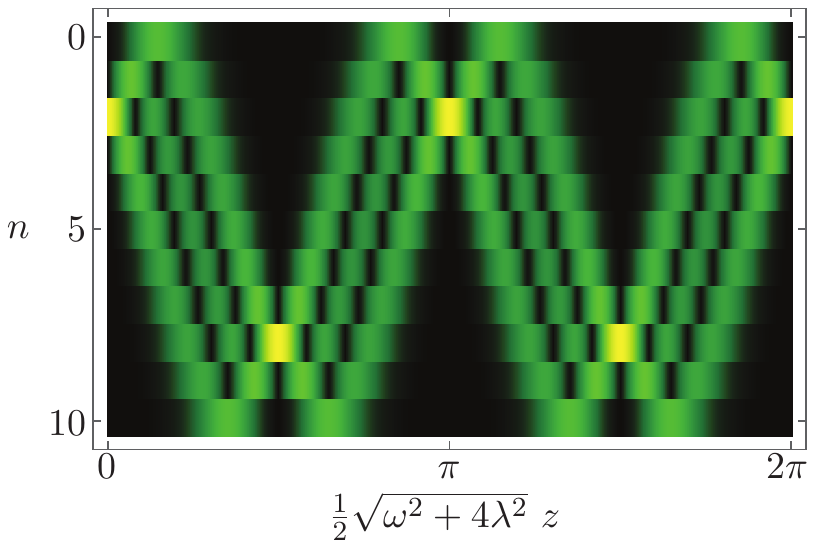}
\caption{Intensity at the $n$-th waveguide, $\vert \mathcal{I}_{m,n}^{(SU(2))}(\omega, \lambda, j, z) \vert^{2}$, for light initially impinging the $m=2$ waveguide of an $SU(2)$ array with parameters $\omega=0$ and $j=5$.}
\label{fig:Fig2}
\end{figure}

In the specific case provided by waveguides with identical effective refractive index, such that the inclusion of an overall constant phase leads to $\omega = 0$, the Hamiltonian-like matrix becomes,
\begin{eqnarray}
\hat{H}_{SU(2)}(0, \lambda, j) &=&  2 \lambda \hat{J}_{x}, \quad \hat{J}_{x} = \frac{1}{2} \left( \hat{J}_{+} + \hat{J}_{-} \right).
\end{eqnarray}
This may be the origin of the $J_{x}$ lattice moniker for such restriction \cite{Perez-Leija2013p022303,PerezLeija2013p012309}. 
In this case it is straightforward to take the limit, 
\begin{eqnarray} 
\lim_{\omega \rightarrow 0} \frac{1}{2} \arctan \frac{2 \lambda}{\omega} = \frac{\pi}{4}
\end{eqnarray}
and use this value to find the normal modes, 
\begin{eqnarray}
\langle n \vert \Omega(m, 0, \lambda,j)_{SU(2)} \rangle = \frac{1}{2^{j}} K_{m}\left(n,\frac{1}{2},2j\right),
\end{eqnarray}
and the impulse function, 
\begin{eqnarray}
\mathcal{I}_{m,n}^{(SU(2))}(0, \lambda, j, z)&=&  \sqrt{ \left( \begin{array}{c} 2j \\ m \end{array}\right) \left( \begin{array}{c} 2j \\ n \end{array}\right)}  \left( i \sin \lambda z \right)^{m+n} \left(\cos \lambda z \right)^{2j-m-n} \times \nonumber \\
&& \times   K_{m}\left(n, \sin^{2} \lambda z , 2j  \right) .
\end{eqnarray}
It is straightforward to notice that the lattice shows a $\pi / \lambda$ periodicity.
It is also simple to check the properties of Krawtchouk polynomials and realize that the initial field amplitude at the $m$-th waveguide will be transferred to the $(2j - m)$-th waveguide  at half the periodicity, $\lambda z = \pi /2$,  
\begin{eqnarray}
\mathcal{I}_{m,n}^{(SU(2))} \left(0, \lambda, j, \frac{\pi}{2 \lambda} \right) = \delta_{n, 2j-m},
\end{eqnarray}
where the notation $\delta_{a,b}$ stands for Kronecker delta that is equal to one when $a=b$ and zero for all other cases.
This has been used for coherent transfer in the so-called $J_{x}$ lattices \cite{Perez-Leija2013p022303,PerezLeija2013p012309} and is an immediate consequence of the impulse function given in terms of trigonometric functions and Krawtchouk polynomials.
Figure \ref{fig:Fig2} shows this coherent transfer and the periodicity discussed above in the intensity, $\vert \mathcal{I}_{m,n}^{(SU(2))}(\omega, \lambda, j, z) \vert^{2}$, of an initial field amplitude impinging just the $m=2$ waveguide of an $SU(2)$ lattice with parameters $\omega=0$ and $j=5$.

\section{Lattices with underlying $SU(1,1)$ symmetry}

There is another three-element closed group that can be used to describe photonic lattices, 
\begin{eqnarray}
\hat{H}_{SU(1,1)}(\omega, \lambda,k) &=& \omega \left( \hat{n} + k \right) + \lambda \left[ \hat{V} \sqrt{ \hat{n} (2k - 1 + \hat{n} ) } +  \sqrt{ \hat{n} (2k - 1 + \hat{n} )  } ~\hat{V}^{\dagger}  \right],\\
&\equiv& \omega \hat{K}_{0} + \lambda ( \hat{K}_{-} + \hat{K}_{+} ).
\end{eqnarray}
Here the operators form the $SU(1,1)$ group that satisfy the commutation relations, $[\hat{K}_{+}, \hat{K}_{-}]= -2 \hat{K}_{0}$ and $[\hat{K}_{0}, \hat{K}^{\pm}]= \pm \hat{K}_{\pm}$.
Here, the ideal lattice has infinite size but these results will describe a real-world finite lattice as long as the propagated electromagnetic field amplitudes stay far from the edge of the array.
The real positive parameter $k>0$ is the Bargmann parameter of the SU(1,1) group \cite{Barut1980}.
Again, this effective Hamiltonian-like matrix can be diagonalized, 
\begin{eqnarray}
\hat{D}_{SU(1,1)}(\xi) \hat{H}_{SU(1,1)}(\omega, \lambda,k) \hat{D}_{SU(1,1)}^{\dagger}(\xi) &=& \sqrt{\omega^{2} - 4 \lambda^{2}} ~\left( \hat{n} + k \right),
\end{eqnarray}
with the difference that a strong restriction appears,
\begin{eqnarray}
\tanh 2 \xi= \frac{2 \lambda}{\omega}, \quad \omega > 2 \lambda.
\end{eqnarray}
Here the displacement operator has the form, 
\begin{eqnarray}
\hat{D}_{SU(1,1)}(\xi) &=& e^{ \xi \hat{K}_{+} - \xi^{\ast} \hat{K}_{-} }, \\
&=& e^{\frac{\xi}{\vert \xi \vert} \tanh \vert \xi \vert \hat{K}_{+}} e^{\ln \mathrm{sech}^{2} \vert \xi \vert \hat{K}_{0}} e^{- \frac{\xi^{\ast}}{\vert \xi \vert} \tanh \vert \xi \vert \hat{K}_{-}},
\end{eqnarray}
where we have used the standard disentangling formulas for $SU(1,1)$ \cite{Wilcox1967p962,Wodkiewicz1985p458,Ban1993p1347}.
Under the restriction, $\omega > 2 \lambda$, the dispersion relation is:
\begin{eqnarray}
\Omega_{SU(1,1)}(m,\omega, \lambda,k) &=& \sqrt{\omega^{2} - 4 \lambda^{2}} ~\left( m + k \right),
\end{eqnarray}
with the $n$-th component of the $m$-th normal states given by the $SU(1,1)$ displaced number states,
\begin{eqnarray}
\langle n \vert \Omega_{SU(1,1)}(m,\omega, \lambda,k) \rangle &=&  \langle n  \vert \hat{D}_{SU(1,1)}(-\xi) \vert m \rangle, \\
&=& \frac{(-1)^n}{\Gamma(2k)} \sqrt{ \frac{\Gamma(m+2k) \Gamma(n+2k)}{m! n!} } \left( \sinh \vert \xi \vert \right)^{m+n} \left( \cosh \vert \xi \vert \right)^{-2k-m-n} \times \nonumber \\
&& \times  ~_{2}F_{1}\left(-m,-n,2k,- \mathrm{csch}^{2} \vert \xi \vert  \right), 
\end{eqnarray}
where the notation $\Gamma(x)$ and $_{2}F_{1}(a,b;c;x)$ stand for the Gamma and Gauss hypergeometric functions, in that order, and we have accounted for the fact that the paremeters $\omega$ and $\lambda$ are real and $\omega > 2 \lambda$.
It is quite important to note that, here, it is only possible to calculate the normal states for the restriction $\omega > 2 \lambda$ with the Gilmore-Perelomov coherent approach due to the characteristics of the $SU(1,1)$ group \cite{Puri2001}.
 
\begin{figure}[ht]
\center\includegraphics[scale=1]{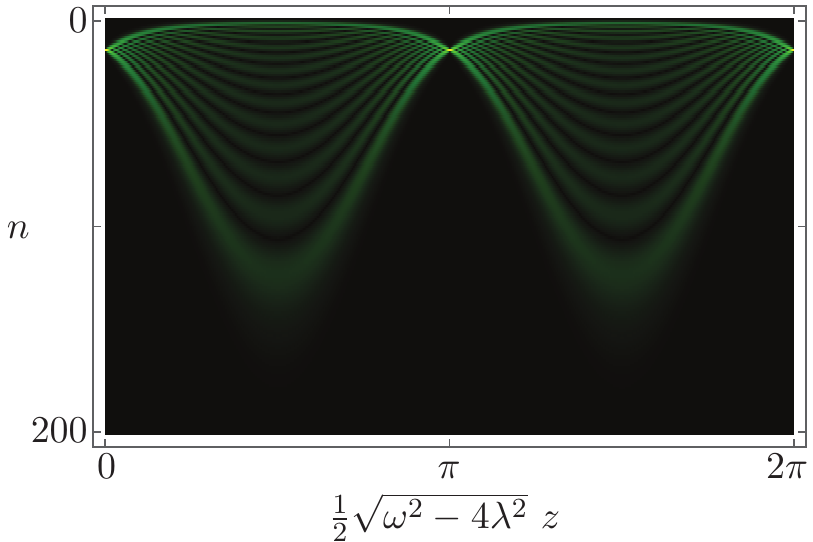}
\caption{Intensity at the $n$-th waveguide, $\vert \mathcal{I}_{m,n}^{(SU(1,1))}(\omega, \lambda, k, z) \vert^{2}$, for light initially impinging the $m=15$ waveguide of a $SU(1,1)$ array with parameters $\lambda = 0.4 \omega$ and $k=1/4$.}
\label{fig:Fig3}
\end{figure}

Again, it is possible to work out an impulse function in terms of what we will call $SU(1,1)$ generalized Gilmore-Perelomov   coherent states,
\begin{eqnarray}
\mathcal{I}_{m,n}^{(SU(1,1))}(\omega, \lambda, k, z) &=&   \langle n  \vert e^{i z \hat{H}_{SU(2)}} \vert m \rangle, \\ &=& \frac{\left(4 \lambda^{2} - \omega^{2} \right)}{\Gamma(2k)} \sqrt{\frac{\Gamma(m+2k) \Gamma(n+2k)}{m! n!}} \left( 2 i \lambda \sinh \frac{z}{2} \sqrt{4 \lambda^{2} - w^{2}} \right)^{m+n} \times \nonumber \\
&& \times \left( \sqrt{4 \lambda^{2} - \omega^{2}} \cosh \frac{z}{2} \sqrt{4 \lambda^{2} - w^{2}} - i \omega \sinh \frac{z}{2} \sqrt{4 \lambda^{2} - w^{2}}  \right)^{-2k-m-n} \times \nonumber \\
&& \times ~_{2}F_{1}\left( -n, -m; 2k; \left(\frac{\omega^{2}- 4\lambda^2}{4\lambda^2}\right) \mathrm{csch}^{2} \frac{z}{2} \sqrt{4 \lambda^{2} - w^{2}} \right).
\end{eqnarray}
Here it is of great importance to note that this impulse function works for any given case of $\omega$ and $\lambda$ as the restriction $\omega > 2 \lambda$ appears only for the dispersion relation and normal states.
Thus, we have two types of propagation regimes. 
The first one is given by the relation $\omega > 2 \lambda$ that transforms the hyperbolic functions into trigonometric functions.
Thus, the lattice will be $2 \pi / \sqrt{\omega^{2} - 2 \lambda^{2}}$ periodical.
Figure \ref{fig:Fig3} shows this periodicity in the squared impulse function, $\vert \mathcal{I}_{m,n}^{(SU(2))}(\omega, \lambda, j, z) \vert^{2}$, of an initial field amplitude impinging just the $m=15$ waveguide of an $SU(1,1)$ lattice with parameters $\lambda=0.4 \omega$ and $k=1/4$.

\begin{figure}[ht]
\center\includegraphics[scale=1]{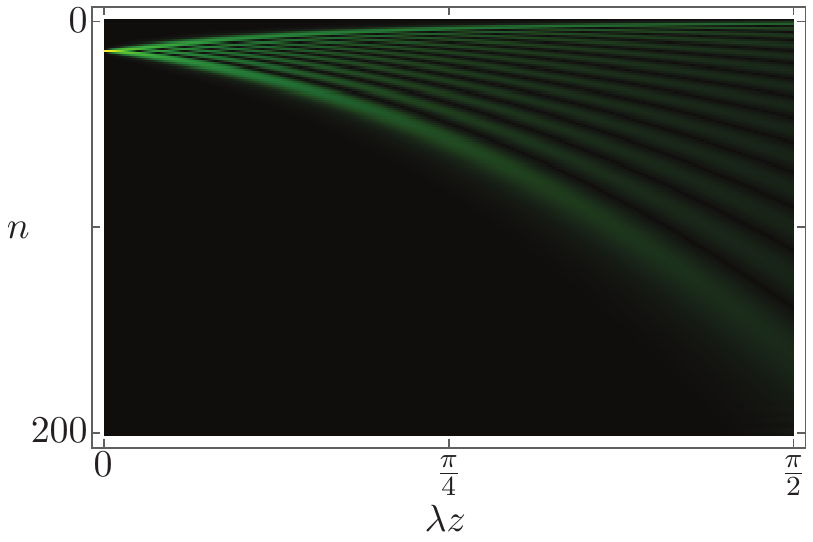}
\caption{Intensity at the $n$-th waveguide, $\vert \mathcal{I}_{m,n}^{(SU(1,1))}(\omega, \lambda, k, z) \vert^{2}$, for light initially impinging the $m=15$ waveguide of a $SU(1,1)$ array with parameters $\omega= 2 \lambda$ and $k=1/4$.}
\label{fig:Fig4}
\end{figure}

\begin{figure}[ht]
\center\includegraphics[scale=1]{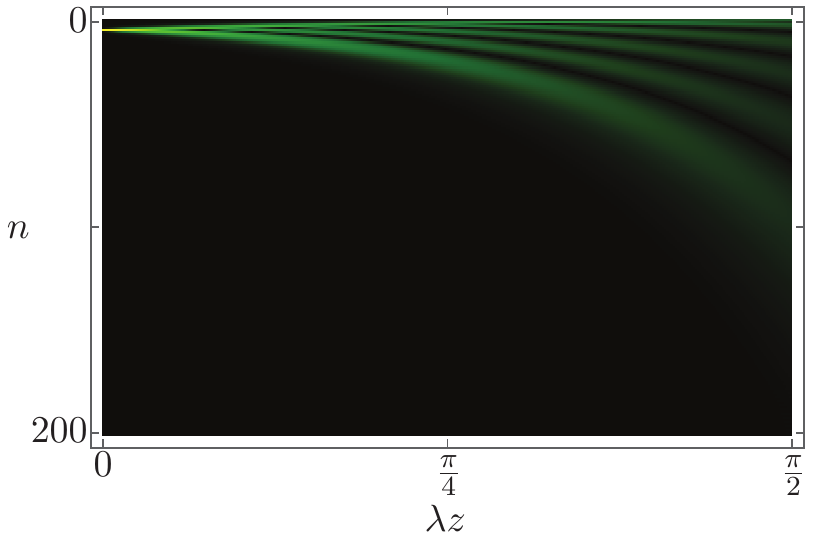}
\caption{Intensity at the $n$-th waveguide, $\vert \mathcal{I}_{m,n}^{(SU(1,1))}(\omega, \lambda, k, z) \vert^{2}$, for light initially impinging the $m=15$ waveguide of a $SU(1,1)$ array with parameters $\omega =0 $ and $k=1/4$.}
\label{fig:Fig5}
\end{figure}

The other will be defined by $\omega \le 2 \lambda$, where the lattice is not periodical and light impinging the $m$-th lattice will disperse with propagation due to the hyperbolic functions. 
In the special cases, $\omega = 2 \lambda$ and $\omega=0$, the impulse function can be calculated via the limits of the hyperbolic functions. 
The impulse function for $\omega = 2 \lambda$, shown in Fig. \ref{fig:Fig4} for Bargmann parameter $k=1/4$, is given by the expression,
\begin{eqnarray}
\mathcal{I}_{m,n}^{(SU(1,1))}(2 \lambda, \lambda, k, z) &=&  \sqrt{\frac{\Gamma(m+2k) \Gamma(n+2k)}{m! n!}} ~\frac{ (-1)^{k+m+n} \left(\lambda z \right)^{m+n} \left(\lambda z + i \right)^{-2k-m-n} }{\Gamma(2k)}  \times \nonumber \\
&& \times ~_{2}F_{1}\left( -n, -m; 2k; - \frac{1}{\lambda^2 z^2 }\right).
\end{eqnarray}
and for $\omega = 0$, shown in Fig. \ref{fig:Fig5} for $k=1/4$, by the expression,
\begin{eqnarray}
\mathcal{I}_{m,n}^{(SU(1,1))}(0, \lambda, k, z) &=&  \sqrt{\frac{\Gamma(m+2k) \Gamma(n+2k)}{m! n!}} ~\frac{\left(  i \sinh \lambda z \right)^{m+n} \left( \cosh \lambda z \right)^{-2k-m-n} }{\Gamma(2k)} \times \nonumber \\
&& \times   ~_{2}F_{1}\left( -n, -m; 2k; - \mathrm{csch}^{2} \lambda z \right).
\end{eqnarray}
Note that a particular realization of lattices with this underlying symmetry described by the parameters $\omega = 1 + \beta^{2}$, $\lambda = \beta$ with the parameter $\beta$ such that $\omega > 2 \lambda$ for the Bargmann parameters $k=1/2$ and $k=1$,  have been used to report pairs of isospectral arrays of photonic lattices as an analogy to supersymmetric quantum mechanics.
In these cases the impulse function reduces to Jacobi polynomials and gave rise to the moniker of Jacobi photonic lattices \cite{ZunigaSegundo2014p987}.

\section{Lattices with underlying Heisenberg-Weyl symmetry}

The Heisenber-Weyl lattice has an effective coupling that increases as the square root of the position of the waveguide, 
\begin{eqnarray}
\hat{H}_{HW}(\omega, \lambda, z) &=& \omega \hat{n} + \lambda \left( \hat{V} \sqrt{\hat{n}} +  \sqrt{\hat{n}} \hat{V}^{\dagger}  \right),\\
&=& \omega \hat{n} + \lambda ( \hat{a} + \hat{a}^{\dagger} ),
\end{eqnarray}
where we have used the creation (annihilation) operators $\hat{a}^{\dagger}$ ($\hat{a}$) that satisfy the commutation relations $[\hat{a}^{\dagger}, \hat{a}]=-1$, $[\hat{n}, \hat{a}^{\dagger}]=\hat{a}^{\dagger}$, and $[\hat{n}, \hat{a}]=-\hat{a}$.
This matrix can be diagonalized,
\begin{eqnarray}
\hat{D}_{HW}(\alpha) \hat{H}_{HW} \hat{D}_{HW}^{\dagger}(\alpha) = \omega \hat{n} - \frac{\lambda^2}{\omega},  \quad \alpha= \frac{\lambda}{\omega}.
\end{eqnarray}
by the Glauber displacement operator \cite{Glauber1963p2766,Sudarshan1963p277},
\begin{eqnarray}
\hat{D}_{HW}(\alpha) = e^{\alpha \hat{a}^{\dagger} - \alpha^{\ast} \hat{a} }.
\end{eqnarray}
Thus, the dispersion relation,
\begin{eqnarray}
\Omega_{HW}(m, \omega, \lambda) &=& \omega m - \frac{\lambda^2}{\omega},
\end{eqnarray}
and the normal modes are given by the displaced Fock states \cite{Satyanarayana85p400,Wunsche1991p359},
\begin{eqnarray}
\langle n \vert \Omega_{HW}(m,\omega, \lambda) \rangle &=&  \langle n \vert \hat{D}_{HW}(-\alpha)\vert m \rangle, \\
&=& \frac{(-1)^{m}  }{\sqrt{m! n!}}   e^{-\frac{1}{2} \vert \frac{\lambda}{\omega}\vert^{2}} \left(-\frac{\lambda}{\omega}\right)^{n-m} U\left( -m,n-m+1, \left\vert \frac{\lambda}{\omega}\right\vert^{2} \right),
\end{eqnarray}
where the $U(a,b,x)$ stands for Tricomi confluent hypergeometric function \cite{Lebedev1965}.
Typically, displaced number states are given in terms of generalized Laguerre polynomilas, but this requires to split the result into two cases, one for $n \ge m$ and another for $ n< m$. 
We favor Tricomi confluent hypergemetric functions because it is straightforward to use them in numeric simulations, rather than defining the cases required with generalized Laguerre polynomials.

\begin{figure}[ht]
\center\includegraphics[scale=1]{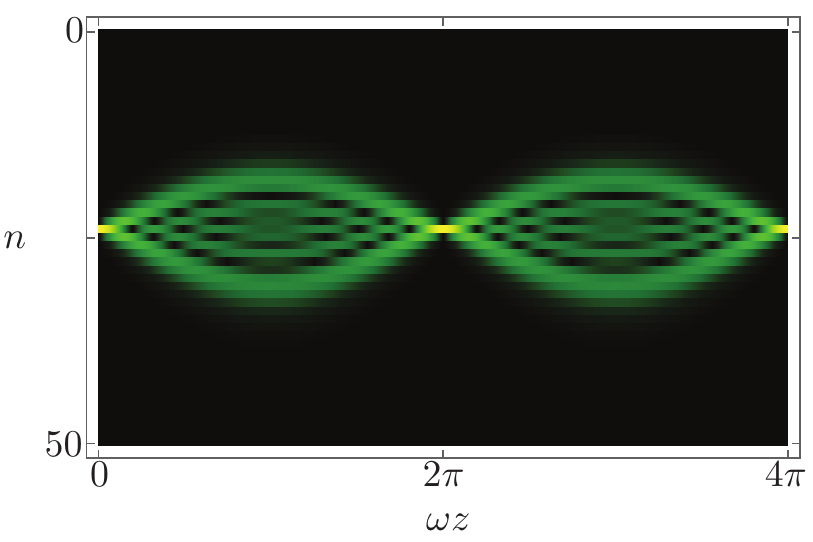}
\caption{Intensity at the $n$-th waveguide, $\vert \mathcal{I}_{m,n}^{(HW)}(\omega, \lambda, z) \vert^{2}$, for light initially impinging the $m=24$ waveguide of a Heisenberg-Weyl array with parameter $\lambda = 0.4 ~\omega $.}
\label{fig:Fig6}
\end{figure}

\begin{figure}[ht]
\center\includegraphics[scale=1]{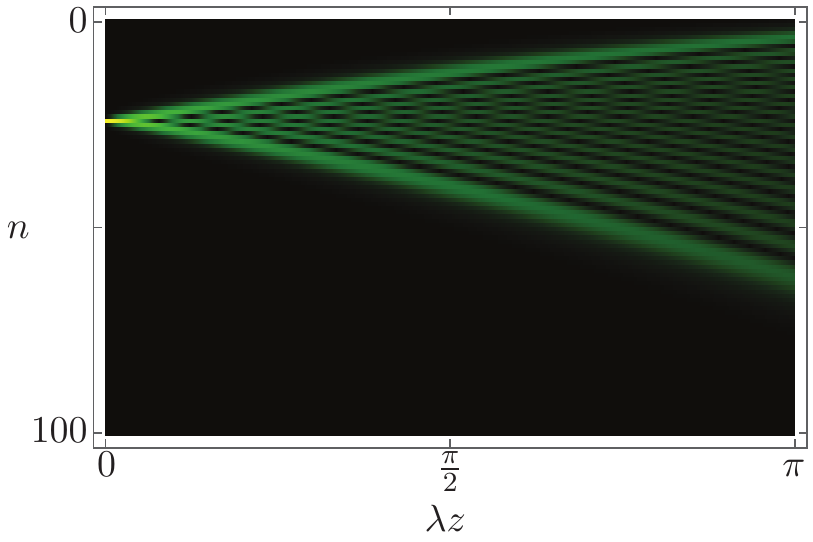}
\caption{Intensity at the $n$-th waveguide, $\vert \mathcal{I}_{m,n}^{(HW)}(\omega, \lambda, z) \vert^{2}$, for light initially impinging the $m=24$ waveguide of a Heisenberg-Weyl array with parameter $\omega=0 $.}
\label{fig:Fig7}
\end{figure}

The impulse function of this array is given by the following expression,
\begin{eqnarray}
\mathcal{I}_{m,n}^{(HW)}(\omega, \lambda, z)&=& \langle n \vert e^{ i z \hat{H}_{HW}} \vert m \rangle, \\
&=& \frac{(-1)^{m}}{\sqrt{m! n!}} e^{ i z \left( m \omega - \frac{\lambda^2}{\omega}  \right) } e^{\vert \frac{\lambda}{\omega} \vert^{2} \left( 1- e^{-i w z} \right)} \left[ \frac{\lambda}{\omega}  \left( e^{i w z} - 1  \right)\right]^{n-m} \times \nonumber \\
&& \times U\left( -m,n-m+1,  4 \vert \frac{\lambda}{\omega} \vert^{2} \sin^{2} \frac{\omega z}{2} \right),
\end{eqnarray}
where we have used the properties of Glauber displacement operator \cite{Klauder1985}.
Thus, light impinging just at just the $m$-th waveguide will produce a field amplitude distribution at the waveguides that is equivalent to the amplitude distribution of the quantum optics displaced number state.
Note that this impulse function has a periodicity of $2 \pi / \omega$ due to the argument in the exponential functions and the $\pi/ \omega$ periodicity of the squared sine argument in Tricomi confluent hypergeometric function.
This is shown in Fig. \ref{fig:Fig6} for an initial field impinging the $m=24$ waveguide in a Heisenberg-Weyl lattice with parameter $\lambda = 0.4 \omega$.
A displaced number state is nothing else than the state obtained by the action of Glauber displacement operator over a Fock state, which may be the origin of the Glauber-Fock lattice moniker given to specific realizations of lattices with underlying Heisenberg-Weyl symmetry \cite{Keil2012p3801,Keil2011p103601,PerezLeija2012p013848,RodriguezLara2011p053845}. 
Note that these well known results for $\omega = 0$ are recovered by taking the limit of the impulse function,
\begin{eqnarray}
\mathcal{I}_{m,n}^{(HW)}(0, \lambda, z) &=& \lim_{\omega \rightarrow 0} \mathcal{I}_{m,n}^{(HW)}(\omega, \lambda,z), \\
&=& \frac{(-1)^m }{\sqrt{m!n!}} e^{ - \frac{1}{2} \vert  \lambda z \vert^{2} }  \left(i  \lambda z \right)^{n-m}  U\left( -m,n-m+1,  \vert \lambda z \vert^{2} \right),
\end{eqnarray}
and in this particular realization the periodicity is lost as shown in Fig. \ref{fig:Fig7}.
A more complicated realization belonging to this symmetry class is provided by what we will call a Lewis-Ermakov lattice, where first and second neighbors couple and effective refractive indices and couplings may include propagation distance dependence \cite{RodriguezLara2014p2083}.

\section{Conclusion}

We have used the well known Gilmore-Perelomov coherent states from quantum mechanics \cite{Gilmore1972p391,Perelomov1972p222,Zhang1990p867} to provide closed form analytic dispersion relations, normal modes and impulse functions for three types of photonic lattices with underlying $SU(2)$, $SU(1,1)$ and Heisenberg-Weyl symmetries.
This symmetry based approach allowed us to analyzed the characteristics of these photonic lattices straight from the impulse functions; in particular, their periodicity or lack of it.
We want to emphasize the fact that the propagator for any given tight-binding array of photonic 
waveguides may be calculated in this way as long as the dynamical group is found, even in the case of arrays were the parameters depend on the propagation distance \cite{RodriguezLara2014p013802,RodriguezLara2014p2083}.
Furthermore, these photonic lattices can be seen as optical simulators of a class of Gilmore-Perelomov generalized coherent states.

\section*{Acknowledgment}

BMRL acknowledges fruitful discussion with Miguel A. Bandres and LVV support from CONACYT master studies grant $\#294880$.

%

\begin{thebibliography}{10}
\newcommand{\enquote}[1]{``#1''}

\bibitem{Longhi2009p243}
S.~Longhi, \enquote{Quantum-optical analogies using photonic structures,} Laser
  Photon. Rev. \textbf{3}, 243 -- 261 (2009).

\bibitem{Longhi2011p453}
S.~Longhi, \enquote{Classical simulation of relativistic quantum mechanics in
  periodic optical structures,} Appl. Phys. B \textbf{104}, 453 -- 468 (2011).

\bibitem{RodriguezLara2015p}
B.~M. Rodr\'iguez-Lara, F.~Soto-Eguibar, and D.~N. Christodoulides,
  \enquote{Quantum optics as a tool for photonic lattice design,} Phys. Scr.
  \textbf{90}, 068014 (2015).

\bibitem{Jones1965p261}
A.~L. Jones, \enquote{Coupling of optical fibers and scattering in fibers,} J.
  Opt. Soc. Am. \textbf{55}, 261 -- 271 (1965).

\bibitem{Snyder1972p1267}
A.~W. Snyder, \enquote{Coupled-mode theory for optical fibers,} J. Opt. Soc.
  Am. \textbf{62}, 1267 -- 1277 (1972).

\bibitem{Kogelnik1972p2327}
H.~Kogelnik and C.~V. Shank, \enquote{Coupled-wave theory of distributed
  feedback lasers,} J. Appl. Phys. \textbf{43}, 2327 -- 2335 (1972).

\bibitem{Yariv1973p919}
A.~Yariv, \enquote{Coupled-mode theory for guided-wave optics,} IEEE J. Quantum
  Elect. \textbf{9}, 919 -- 933 (1973).

\bibitem{Streifer1987p1}
W.~Streifer, M.~Osi\'nski, and A.~Hardy, \enquote{Reformulation of the
  coupled-mode theory of mutiwaveguide systems,} J. Lightwave Technol.
  \textbf{5}, 1 -- 4 (1987).

\bibitem{Haus1987p16}
H.~A. Haus, W.~P. Huang, S.~Kawakami, and N.~A. Whitaker, \enquote{Coupled-mode
  theory of optical waveguides,} J. Lightwave Technol. \textbf{5}, 16 -- 23
  (1987).

\bibitem{Huang1994p963}
W.-P. Huang, \enquote{Coupled-mode theory for optical waveguides: an overview,}
  J. Opt. Soc. Am. A \textbf{11}, 963 -- 983 (1994).

\bibitem{RodriguezLara2014p2083}
B.~M. Rodr{\'\i}guez-Lara, P.~Aleahmad, H.~M. Moya-Cessa, and D.~N.
  Christodoulides, \enquote{Ermakov-lewis symmetry in photonic lattices,} Opt.
  Lett. \textbf{39}, 2083--2085 (2014).

\bibitem{RodriguezLara2014p013802}
B.~M. Rodr{\'\i}guez-Lara, H.~M. Moya-Cessa, and D.~N. Christodoulides,
  \enquote{Propagation and perfect transmission in three-waveguide axially
  varying couplers,} Phys. Rev. A \textbf{89}, 013802 (2014).

\bibitem{Wilcox1967p962}
R.~M. Wilcox, \enquote{Exponential operators and parameter differentiation in
  quantum physics,} J. Math. Phys. \textbf{8}, 962 -- 982 (1967).

\bibitem{Wodkiewicz1985p458}
K.~W\'odkiewicz and J.~H. Eberly, \enquote{Coherent states, squeezed
  fluctuations, and the $SU(2)$ and $SU(1,1)$ groups in quantum-optics
  applications,} J. Opt. Soc. Am. B \textbf{2}, 458 -- 466 (1985).

\bibitem{Ban1993p1347}
M.~Ban, \enquote{Decomposition formulas for $su(1,1)$ and $su(2,2)$ Lie
  algebras and their applications in quantum optics,} J. Opt. Soc. Am. B
  \textbf{10}, 1347 -- 1359 (1993).

\bibitem{Gilmore1972p391}
R.~Gilmore, \enquote{Geometry of symmetrized states,} Ann. Phys. \textbf{74},
  391--463 (1972).

\bibitem{Perelomov1972p222}
A.~M. Perelomov, \enquote{Coherent states for arbitrary Lie group,} Comm. Math.
  Phys. \textbf{26}, 222 -- 236 (1972).

\bibitem{Zhang1990p867}
W.~M. Zhang, D.~H. Feng, and R.~Gilmore, \enquote{Coherent states: Theory and
  some applications,} Rev. Mod. Phys. \textbf{62}, 867 (1990).

\bibitem{PerezLeija2013p012309}
A.~Perez-Leija, R.~Keil, A.~Kay, H.~Moya-Cessa, S.~Nolte, L.-C. Kwek,
  B.~Rodr\'{i}guez-Lara, A.~Szameit, and D.~Christodoulides, \enquote{Coherent
  quantum transport in photonic lattices,} Phys. Rev. A \textbf{87}, 012309
  (2013).

\bibitem{Perez-Leija2013p022303}
A.~Perez-Leija, R.~Keil, H.~Moya-Cessa, A.~Szameit, and D.~N. Christodoulides,
  \enquote{Perfect transfer of path-entangled photons in $J_{x}$ photonic
  lattices,} Phys. Rev. A \textbf{87}, 022303 (2013).

\bibitem{ZunigaSegundo2014p987}
A.~Zu\~niga-Segundo, B.~M. Rodr\'iguez-Lara, D.~J.~Fern\'andez C., and H.~M. Moya-Cessa,
  \enquote{Jacobi photonic lattices and their SUSY partners,} Opt. Express
  \textbf{22}, 987 -- 994 (2014).

\bibitem{PerezLeija2010p2409}
A.~Perez-Leija, H.~Moya-Cessa, A.~Szameit, and D.~N. Christodoulides,
  \enquote{Glauber-Fock photonic lattices,} Opt. Lett. \textbf{35}, 2409 --
  2411 (2010).

\bibitem{Keil2011p103601}
R.~Keil, A.~Perez-Leija, F.~Dreisow, M.~Heinrich, H.~Moya-Cessa, S.~Nolte,
  D.~N. Christodoulides, and A.~Szameit, \enquote{Classical analogue of
  displaced Fock states and quantum correlations in Glauber-Fock photonic
  lattices,} Phys. Rev. Lett. \textbf{107}, 103601 (2011).

\bibitem{RodriguezLara2011p053845}
B.~M. Rodr{\'\i}guez-Lara, \enquote{Exact dynamics of finite Glauber-Fock
  photonic lattices,} Phys. Rev. A \textbf{84}, 053845 (2011).

\bibitem{Keil2012p3801}
R.~Keil, A.~Perez-Leija, P.~Aleahmad, H.~Moya-Cessa, S.~Nolte, D.~N.
  Christodoulides, and A.~Szameit, \enquote{Observation of Bloch-like revivals
  in semi-infinite Glauber-Fock photonic lattices,} Opt. Lett. \textbf{37},
  3801 -- 3803 (2012).

\bibitem{PerezLeija2012p013848}
A.~Perez-Leija, R.~Keil, A.~Szameit, A.~F. Abouraddy, H.~Moya-Cessa, and D.~N.
  Christodoulides, \enquote{Tailoring the correlation and anticorrelation
  behavior of path-entangled photons in Glauber-Fock lattices,} Phys. Rev. A
  \textbf{85}, 013848 (2012).

\bibitem{Barut1980}
A.~O. Barut and R.~Raczka, \emph{Theory of group representations and
  applications} (PWN Polish Scientific Publishers, 1980).

\bibitem{Askey1975}
R.~Askey, \emph{Orthogonal Polynomials and special functions}, CBMS-NSF
  Regional Conference Series in Applied Mathematics (SIAM, 1975).

\bibitem{Wolf2007p651}
K.~B. Wolf and G.~Kr\"otzsch, \enquote{Geometry and dynamics in the fractional
  discrete Fourier transform,} J. Opt. Soc. Am. A \textbf{24}, 651 -- 658
  (2007).

\bibitem{Vicent2008p1875}
L.~E. Vicent and K.~B. Wolf, \enquote{Unitary transformation between Cartesian-
  and polar-pixellated screens,} J. Opt. Soc. Am. A \textbf{25}, 1875 -- 1884
  (2008).

\bibitem{Wolf2009p509}
K.~B. Wolf, \enquote{Mode analysis and signal restoration with Kravchuk
  functions,} J. Opt. Soc. Am. A \textbf{26}, 509 -- 516 (2009).

\bibitem{Vicent2011p808}
L.~E. Vicent and K.~B. Wolf, \enquote{Analysis of digital images into
  energy-angular momentum modes,} J. Opt. Soc. Am. A \textbf{28}, 808 -- 814
  (2011).

\bibitem{Puri2001}
R.~R. Puri, \emph{Mathematical methods of quantum optics} (Springer, 2001).

\bibitem{Glauber1963p2766}
R.~J. Glauber, \enquote{Coherent and incoherent states of the radiation field,}
  Phys. Rev. \textbf{131}, 2766 -- 2788 (1963).

\bibitem{Sudarshan1963p277}
E.~C.~G. Sudarshan, \enquote{Equivalence of semiclassical and quantum
  mechanical descriptions of statistical light beams,} Phys. Rev. Lett.
  \textbf{10}, 277 -- 279 (1963).

\bibitem{Satyanarayana85p400}
M.~V. Satyanarayana, \enquote{Generalized coherent states and generalized
  squeezed coherent states,} Phys. Rev. D \textbf{32}, 400 -- 404 (1985).

\bibitem{Wunsche1991p359}
A.~W{\"u}nsche, \enquote{Displaced Fock states and their connection to
  quasiprobabilities,} Quantum Opt. \textbf{3}, 359 -- 383 (1991).

\bibitem{Lebedev1965}
N.~N. Lebedev, \emph{Special functions and their applications} (Prentice-Hall, 1965).

\bibitem{Klauder1985}
J.~R. Klauder and B.-S. Skagerstam, \emph{Coherent states. Applications in
  physics and mathematical physics} (World Scientific, 1985).

\end{thebibliography}

\end{document}